\documentclass[12pt]{article}
\usepackage{euscript}

\usepackage{latexsym}
\usepackage{amsmath, amssymb,graphics}
\usepackage[all]{xy}

\usepackage{color}

\usepackage{fullpage}

\renewcommand{\(}{\begin{equation*}}
\renewcommand{\)}{\end{equation*}}
\newcommand{\bea}{\begin{eqnarray*}}
\newcommand{\eea}{\end{eqnarray*}}
\newcommand{\R}{{\mathbb R}}

\newcommand{\Z}{{\mathbb Z}}

\newcommand{\T}{{\mathbb T}}

\def\vol{{\rm vol}}

\newcommand{\cG}{\ensuremath{\mathcal G}}

\newcommand{\cM}{\ensuremath{\mathcal M}}

\newcommand{\bo}{\raise-1mm\hbox{\Large$\Box$}}              % D'Alembertian
\newcommand{\beq}{\begin{equation}}
\newcommand{\eeq}{\end{equation}}

\numberwithin{equation}{section}

\renewcommand{\(}{\begin{equation}}
\renewcommand{\)}{\end{equation}}

\newcommand{\RR}{{\mathbb R}}
\newcommand{\ZZ}{{\mathbb Z}}

\def\R{{\mathbb R}}
\def\Z{{\mathbb Z}}

\def\1{{\bf 1}}

\def\<{\langle}
\def\>{\rangle}

\numberwithin{equation}{section}

\renewcommand{\(}{\begin{equation}}
\renewcommand{\)}{\end{equation}}
\newbox\ncintdbox \newbox\ncinttbox
\setbox0=\hbox{$-$} \setbox2=\hbox{$\displaystyle\int$}
\setbox\ncintdbox=\hbox{\rlap{\hbox to
\wd2{\hskip-.125em\box2\relax \hfil}}\box0\kern.1em}
\setbox0=\hbox{$\vcenter{\hrule width 4pt}$}
\setbox2=\hbox{$\textstyle\int$}
\setbox\ncinttbox=\hbox{\rlap{\hbox to
\wd2{\hskip-.175em\box2\relax \hfil}}\box0\kern.1em}
\newcommand{\ncint}{\mathop{\mathchoice{\copy\ncintdbox}%
{\copy\ncinttbox}{\copy\ncinttbox}{\copy\ncinttbox}}\nolimits}

\begin{document}

\begin{titlepage}
%\begin{flushright}

%hep-th/xxxxxxx
%\end{flushright}

\vspace{2em}
\def\thefootnote{\fnsymbol{footnote}}

\begin{center}
{\Large\bf  
Higher abelian gauge theory associated to gerbes on\\ noncommutative deformed 
M5-branes and S-duality
%M5-branes
}
\end{center}
\vspace{3em}

\begin{center}
Varghese Mathai\footnote{e-mail: {\tt
mathai.varghese@adelaide.edu.au}. Department of Pure Mathematics, University of Adelaide, Adelaide 5005, Australia.
}
and 
Hisham Sati\footnote{e-mail: {\tt
hsati@pitt.edu}. Department of Mathematics, University of Pittsburgh, Pittsburgh, PA 15260, USA.
}
\end{center}

\vspace{3em}
\begin{abstract}
We enhance the action of higher abelian gauge theory associated to a gerbe on 
an M5-brane with an action of a torus $\T^n (n\ge 2)$, by a noncommutative $\T^n$-deformation of 
the M5-brane. The  ingredients of the noncommutative action and equations of motion  
 include the  deformed  Hodge duality, deformed wedge product, and the noncommutative
 integral over the noncommutative space obtained by strict deformation quantization. 
As an application we then introduce a variant model with an enhanced action in which we show that 
the corresponding partition function is a modular form, which is a purely noncommutative 
geometry phenomenon since the usual theory only has a $\mathbb Z_2$-symmetry. 
In particular, S-duality 
in this 6-dimensional higher abelian gauge theory model is shown to be, in this sense,  
on par with the usual 4-dimensional case.

 \end{abstract}

\end{titlepage}

\tableofcontents

%%%%%%%%%
\section{Introduction}
%%%%%%%%

The presence of a B-field in string theory generally makes the underlying space 
noncommutative (see \cite{SW}). This can appear on the worldvolume theory of branes as well 
as in space-time. D-branes support Yang-Mills fields, so the presence 
of the B-field leads to noncommutative Yang-Mills theory (see \cite{CR}). 
Recently there has been a lot of interest in the theory on the worldvolume of the
M-theory fivebrane (M5-brane). This is a superconformal field theory (in some limit) which 
has a B-field in its field content (see e.g. \cite{B} for a survey). 
For a special case of the worldvolume theory of the fivebrane, 
 the ADHM construction can be 
extended to describe (effectively 4-dimensional) noncommutative instantons  \cite{NS}. 
 We will generalize an aspect of this to consider the full six-dimensional theory 
 without relying on the reduction to Yang-Mills theory, although we will
 also consider interesting instances of such a reduction.  
  As the fivebrane theory is believed to be intrinsically quantum, a description in terms of 
noncommutative geometry would be appropriate. Indeed, this has been considered previously 
in \cite{BBSS} using the C-field on the worldvolume to describe noncommutative 
fivebranes via open membranes. 
 We will instead make use of the notion of strict deformation quantization of 
 Rieffel 
 %\cite{R1} \cite{R2} 
 \cite{R4}. 
  For a parametrized version of strict deformation quantization and its applications
 to T-duality, see \cite{HM}.

\vspace{3mm}
The theory on the worldvolume of a single  fivebrane, given by a closed oriented 6-dimensional manifold,
can be described as an abelian gerbe theory (cf. \cite{Bry,BCMMS,MS} for extensive description of gerbes).
On the other hand, an action principle for gerbes
was studied in \cite{MR} 
\(
S(B)= \int_{M^6} H_B \wedge * H_B\;,
\label{abelianYM}
\)
where $B$ is the B-field of the gerbe whose 3-curvature is $H_B=dB$ with Dixmier-Douady class
 $[H_B]\in H^3(M;\ZZ)$. This theory only has a $\ZZ_2$-symmetry given by $B \to -B$.
  Such an action at the level of differential forms also appears in the literature as describing 
part of the dynamics of the fivebrane, but would vanish upon imposing the desired self-duality 
equations. We will take this as the starting point to propose, with a remedy, 
an action for the fivebrane including noncommutativity and self-duality. 
%Even though there is no action for the self-dual 
%3-form, there is a quantum field theory
%of closed selfdual H-field. 
Classically, the equations describing the dynamics of the H-field are
\(
d^\dagger H_B=0\;, \qquad dH_B=0\;,
\label{EOM}
\)
where $*$ is the Hodge duality operator in six dimensions, $d$ is the exterior 
derivative on forms, and $d^\dagger$ is the adjoint of $d$.  The solutions are the harmonic 3-forms on 
the M5-brane $M^6$. One can also add source terms and also impose self-duality 
by hand, in which case the action \eqref{abelianYM} might be referred to 
as a pseudo-action (see \cite{BM}). Recent accounts of higher abelian gauge theory 
in this context, via differential cohomology, are given in \cite{FSS1, FSS2, FSS3, Sz}.

\medskip
Note that in our setting the source of commutativity will not be a physical parameter 
 like the C-field, 
 but rather a mathematical parameter that we introduce into the theory. As such, 
 from the physical point 
 of view our setting describes a
 a six-dimensional  model of abelian gerbe theory associated to the M5-brane. 
At this stage it might benefit the reader to dissect the title of this paper: The model that 
we study is the ``Higher abelian gauge theory" and we have explained above that it is 
one that essentially appears on the worldvolume of M5-branes. Furthermore, 
we con side the underlying M5-brane worldvolumes to be ``noncommutative deformed"
by adding a parameter $\theta$ via Rieffel's strict deformation quantization. Aside from studying the 
model for its own right, we also discover that there are certain desirable byproducts of doing so.
The first is that one can study ``S-duality" in this setting in essentially a similar way that 
is done for usual S-duality in four dimensions. The latter has far-reaching consequences 
on geometric topology and the Geometric Langlands Program \cite{W-6, W-knot}.
We hope that this paper is a first step in extending these connections directly 
to six dimensions.

%%%%%%%%%%%
\paragraph{The main content.} We provide the following fix to the action whenever
the M5-brane $M^6$ has an action of a torus $\T^n$ for some $n\ge 2$. 
To give the action and the partition function (of a variant model) we describe 
a noncommutative $\T^n$-deformation $M_\theta^6$ of the M5-brane worldvolume, 
of the integral, 
of the Hodge star operator, and of the wedge product. 
We propose the new action, where we will assume that 
3-curvature is self-adjoint $H_B^\dagger = H_B$ with respect to the deformed inner product,
\(
S_\theta(B)= \int_{M_\theta} H_B \wedge_\theta *_\theta H_B 
- \int_{M_\theta^6} H_B \wedge_\theta  H_B +
\int_{M_\theta^6} C_6\;,
\label{Eq action1}
\) 
and where $C_6$ is a potential for the dual of the C-field in M-theory 
 pulled back to the fiverbrane worldvolume.

\medskip
 The equations of motion (EOM) and Bianchi identity
 \footnote{For brevity, we will refer to both as EOMs.}
   of the H-field can be derived as in the case of 
 noncommutative Yang-Mills \cite{LS} to be
\(
d^{\dagger_\theta} H_B=0, \qquad dH_B=0\;,
\label{EOM2}
\)
where $d^{\dagger_\theta}$ is the adjoint of $d$ with respect to the deformed inner product
$$
(H, H')_\theta = \int_{M^6_\theta} H^\dagger \wedge_\theta *_\theta H' .
$$ 
Since $d^{\dagger_\theta} = \pm *_\theta d *_\theta$ (cf. \cite{LS}), the EOM in \eqref{EOM2} can be re-written as 
\(
d*_\theta H_B=0, \qquad dH_B=0\;.
\label{EOM3}
\)

\noindent {\bf 1.} The main goal is to show that the above action makes sense by describing the following 
ingredients in the next section,\\

\noindent {\bf (a)} the noncommutative space $M_\theta^6$, as a deformation of 
the worldvolume $M^6$, 

\noindent {\bf (b)} the operators corresponding to $H_3$ and 
$C_6$ at the quantum level,  

\vspace{1mm}
\noindent {\bf (c)} the deformed wedge product $\wedge_\theta$,

\vspace{1mm}
\noindent {\bf (d)} the deformed Hodge star operator
$*_\theta$, and 

\vspace{1mm}
\noindent {\bf (e)} the noncommutative integral $\displaystyle\int_{M_\theta^6}$. 

\vspace{1mm}
\noindent 
We hope these will be of independent interest for other settings as well. 
Such deformations have the virtue of keeping key properties of the underlying classical 
geometry unaltered, and as such many classical concepts can be extended
through the deformation in this approach.

\medskip
\noindent {\bf 2.} The second goal is setting up the partition function, which we 
do in Sec. \ref{Sec pf}.
In the commutative case, self-duality of $(2p+1)$-forms in $4p+2$ dimensions
poses a problem as far as an action principle goes. Naively, the kinetic term 
would vanish identically upon  imposing self-duality, as the action would then be 
the wedge product of an odd degree differential form with itself.
Consequently, this then implies a problem for the partition function. 
We would like to warn the reader that the fivebrane partition 
function is a complicated issue due to the self-duality constraint and that the usual method of 
dealing with it is via holomorphic factorization, which leads to an ambiguityy in its definition and
 is related to Spin structures on the worldvolume (see \cite{Wi} for 
a detailed explanation).
There have been proposals to 
evade this by not working directly in $4p+2$ dimensions but rather extending to 
a Chern-Simons theory in $4p+3$ dimensions and/or to the bounding 
theory in $4p+4$ dimensions. This allows the partition function to be defined
as a section of a line bundle over the intermediate Jacobian, and requires
a quadratic refinement \cite{Wi}. Discussions on extension to (higher) differential 
cohomology and stacks are given in \cite{FSS1, FSS2, FSS3}. 
The formulation that we propose via noncommutative geometry does not
suffer from such an immediate problem, because ultimately 
$H_{2p+1} \wedge_\theta H_{2p+1}\neq 0$. Therefore, for $p=1$, the noncommutative
deformation removes this subtlety or difficulty that has plagued the 
study of the M5-brane.
\footnote{An analogous argument  extends to the other physically 
important cases, namely the self-dual scalar 
in $d=2$ and to type IIB string theory in $d=10$, i.e. for $p=0$ and
$p=2$, respectively.}

\vspace{3mm}
\noindent {\bf 3.} The third goal is to calculate the partition function of the model. 
For this we follow the discussion of the 4-dimensional abelian Yang-Mills case 
to evaluate the partition function via functional determinants. This involves summing over
gerbe connections and requires an extension of some analytic concepts, such as Hodge 
theory, from the classical to the noncommutative (in our sense) setting. This leads to 
determinants of Laplace-type operators, which we interpret via Faddeev-Popov 
ghosts as well as ghost-for-ghost determinants. A regularization of the determinants 
is needed, and we choose the $\zeta$-function regularization. 
 Extension to the global 
case when the gerbe is not trivial requires summing over the moduli spaces of 
solutions to the gerbe curvature equations. We also account for torsion 
in the curvature, which extends the moduli space of solutions (instantons).

\vspace{3mm}
\noindent {\bf 4.} 
The fourth goal, which can be viewed as an application of the above construction,  
is to provide a variant model in which we can study a form of S-duality. This 
model, in which we explore how much certain aspects of the 4-dimensional undeformed theory 
can be carried over to the 6-dimensional deformed theory, will not 
fully capture the dynamics of the M5-brane theory, but we hope it will nevertheless give some insight into that theory.
We modify the action \eqref{Eq action1} to include coupling parameters 
\footnote{We emphasize the fact that this where we depart from the M5-brane theory, which does not have
adjustable coupling parameters. We thank David Berman for remarks on this point.}
%\(
%S_\theta(B; \tau)= \frac{4\pi }{e^2}\int_{M_\theta} H_B \wedge_\theta *_\theta H_B 
%-\frac{i\Theta}{2\pi} \int_{M_\theta^6} H_B \wedge_\theta  H_B +
%\int_{M_\theta^6} C_6\;,
%\
%\) 
\(
S_\theta(B; \tau)= \frac{1}{2e^2}\int_{M_\theta^6} H_B \wedge_\theta *_\theta H_B + 
\frac{i \Theta}{2} \int_{M_\theta^6} H_B \wedge_\theta  H_B + \int_{M_\theta^6} C_6\;,
\label{Eq action}
\)
with coupling parameter $\tau= \frac{\Theta}{2\pi} + \frac{4\pi i}{e^2}$,
The main point to highlight 
is that the deformed wedge product is no longer skew-symmetric,
so that now it is possible to restore the full
$SL(2, \Z)$ symmetry to the partition function associated to the action $S_\theta(B; \tau)$ and, furthermore, 
that it is a purely noncommutative geometry phenomenon, similar in spirit to the  renowned work by 
Nekrasov and Schwartz \cite{NS} in another context as alluded to earlier.

\medskip
We also study the partition  function $Z_\theta(M_\theta^6; \tau)$ associated to the action  $S_\theta(B; \tau)$
and show that it is a modular form, and identify the modular weights. Furthermore, under some assumptions on 
heat kernels in the noncommutative setting, we also discuss curvature corrections to these modular weights. 
A full account would require an extension of the seminal work of Vafa-Witten \cite{VW} to our
setting, which goes far beyond the scope of the current paper. Nevertheless, we expect this fact to have 
important consequences 
somewhat similar to the far-reaching 
4-dimensional case. Traditionally, modularity in four dimensions is explained by dimensional 
reduction on the 
%2-torus from a six dimensional theory, which generally does any possess 
%any modularity. What 
2-torus from a six dimensional theory, which generally does not possess any modularity. What
we do here uncovers modularity already in six dimensions, a striking 
phenomenon that is a result of noncommutativity there.  \\

\noindent{\bf Acknowledgements.} The authors are grateful to Keith Hannabuss for careful reading of the manuscript
and for useful feedback and to David Berman for valuable and crucial comments on the first version. 
The author are also grateful to the two anonymous referees for very useful suggestions, both on the 
mathematical and the physical sides, that 
greatly improved the paper from its original form. 
V.M. thanks the 
Australian Research Council for support via ARC Discovery Project grants 
DP110100072 and DP130103924.
H.S. thanks the National Science Foundation for support via NSF Grant PHY-1102218.

%%%%%%%%%%%%
\section{The noncommutative setting for the fivebrane}
%%%%%%%%%%

In this section we provide the main construction and proposal of this letter, which is 
to enhance aspects of the fivebrane worldvolume theory to the noncommutative 
 setting via strict deformation quantization. The virtue of this approach is that 
 the ingredients and calculations are relatively transparent hence 
 utilizable in calculations and, furthermore, can be adapted to other settings. 
 A good part of the ensuing discussion is an adaptation to our setting 
 of known constructions in noncommutative geometry, but there are also 
 new constructions and definitions; in particular, the noncommutative wedge
 product in Sec. \ref{NC ^}.
 
%%%%%%%%%
\subsection{Noncommutative worldvolumes}
%%%%%%%%%

 Let $M^6$ be a compact Riemannian Spin manifold (without boundary)
 of dimension six
whose isometry group has rank $r \geq 2$. Then $M^6$ admits natural isospectral deformations 
to noncommutative geometries $M_\theta$, with an antisymmetric deformation parameter
$\theta=(\theta_{ab}=-\theta_{ba})$, $\theta_{ab}\in \R$;
see \cite{CD, CL, LS}, which we follow in this section.
The idea is to deform the standard spectral triple describing the Riemannian geometry of 
$M^6$ along a torus embedded in the isometry group to get an isospectral triple 
$(C^\infty (M_\theta), \cal{H}, D, \gamma)$, where $\cal{H}$ is the Hilbert space, $D$ is 
the Dirac operator, and $\gamma$ is the chirality operator
\cite{CL}. This is done by deforming the torus action.

\vspace{3mm}
The natural one-parameter deformation can be taken to be isospectral, 
%i.e. unchanging the Dirac operator $D$, and the algebra
i.e. leaving the Dirac operator $D$ unchanged,  and the algebra 
of smooth functions $C^\infty (M^6_\theta)$ in the 
noncommutative geometry $M_\theta^6$ can be described in terms of 
the quantization of smooth functions $L_\theta (C^\infty (M^6))$
on the underlying classical geometry $M^6$. 
The noncommutative Spin geometry will be 
 $(L_\theta(C^\infty (M)), \cal{H}, D)$. 
%The spectrum of the operator is the same as that of the 
%Dirac operator $D$ on $M$,
Note that in this approach
 all spectral properties are preserved.

\vspace{3mm}
Consider the isometric smooth action 
$\sigma$ of $\T^n$, $2 \leq n \leq 6$, on $M^6$. 
Decompose the classical algebra of smooth functions $C^\infty (M^6)$ 
into spectral subspaces indexed by the dual group $\Z^n=\widehat{\T}^n$: 
each $r \in \Z^n$ labels a character of $\T^n$ via $e^{2\pi i s} \mapsto e^{2\pi i r\cdot s}$.
The $r$-th spectral subspace for $\sigma$ on $C^\infty (M^6)$ is formed of functions $f_r$ such that 
$\sigma_s(f_r)=e^{2\pi i r \cdot s}f_r$, each $f\in C^\infty (M^6)$ is the sum 
of a unique (rapidly convergent) series $f=\sum_{r \in \Z^n} f_r$.
%\paragraph{The $\theta$-deformation of $C^\infty (M^6)$.} 
With  $\theta=(\theta_{jk}=-\theta_{kj})$ a real antisymmetric $n \times n$ matrix,
replace the ordinary product by a deformed product
$
f_r \times_\theta g_{r'}:= f_r \sigma_{\frac{1}{2}r \cdot \theta} (g_{r'})=e^{\pi i r \cdot \theta \cdot r'} f_r g_{r'}$,
and denote $C^\infty(M^6_\theta):= (C^\infty (M^6), \times_\theta)$. 
The action $\sigma$ of $\T^n$ extends to $C^\infty (M^6_\theta)$.
At the level of the $C^*$-algebra of continuous functions one has a strict deformation quantization in the 
direction of the Poisson structure defined by the matrix $\theta$. 
The quantization of smooth functions is given by the {\em quantization map}
\(
L_\theta: C^\infty (M^6) \to C^\infty (M^6_\theta)\;,
\)
 which satisfies 
$
L_\theta (f \times_\theta g)= L_\theta (f) L_\theta (g)$. See \cite{CD, CL, LS} for more details.

%%%%%%%%%%
\subsection{The noncommutative integral}
%%%%%%%%%%
Corresponding to the 
spectral triple $(C^\infty (M_\theta), {\cal{H}}, D)$
%is $6^+$-summable and there 
is a noncommutative 
integral defined as a Dixmier trace (see \cite{C})
\(
\ncint L_\theta (f) := {\rm Tr}_\omega (L_\theta (f) |D|^{-6}) 
\)
with $f \in C^\infty (M^6)$ via its representation on
the Hilbert space $\cal{H}$.

\paragraph{The $C_6$-integral.} 
We will use the following \cite{GIV} \cite{LS} as the definition of the volume 
form on $M_\theta^6$
\(\label{ncint}
\ncint L_\theta (f) = \int_{M^6} f d\nu\;. 
\)
The integral over $M_\theta^6$ can be defined using the quantum integral of 
the operators corresponding to the differential forms. For $C_6\in \Omega^6(M_\theta^6)$ 
we define
\(
\int_{M_\theta^6} C_6:=\ncint *_\theta C_6\;,
\)
where $*_\theta C_6$ is an element in $C^\infty (M_\theta^6)$ and the right-hand side
 is defined as in \eqref{ncint}.
If $C_6$ is an exact form, that is if $C_6=dA_5$ for some 
5-form $A_5 \in \Omega^5(M_\theta^6)$, then it can be checked that
the integral of $C_6$ will be zero.
%$
%\int_{M_\theta^6} C_6=\int_{M_\theta^6} dA_5=0$, since the integral can be written as 
%$\ncint d_D A_5^D= \ncint d_D L_\theta (A_5^{D (0)})$,
%where $A_5^{D (0)}$ is the classical counterpart of $A_5$, that is 
%$A_5= L_\theta (A_5^{D (0)})$. This reduces to a classical integral 
%which vanishes (see Lemma 6 in \cite{LS}). 
This is consistent with -- and is in a sense a quantum version of --
the usual requirement of having 
fivebranes with no boundaries. \\

The other integrals in the action are defined in an analogous manner.

%%%%%%%%%%%%
\subsection{Deformed wedge product}
%%%%%%%%%%%%
\label{NC ^}

Consider the action of $\T^n, \, n\ge 2,$ on $M^6$. This action induces an 
action of $\T^n$ on the space of differential forms $\Omega (M^6)$ on $M^6$. 
Starting with a $U(1)$-cocycle $\theta \in Z^2(\widehat{\T^n}, U(1))$ on the dual to the torus 
$\widehat{\T^n}$,
 we would like to 
deform the wedge product. To that end, we 
decompose the space of differential forms with respect to the 
characters of the torus group
\(
\Omega(M^6) \cong \bigoplus_{\alpha \in \widehat{\T^n}} \Omega (M^6)_\alpha\;,
\label{Form dec}
\)
where the components in the decomposition are given by
\(
\Omega (M^6)_\alpha:= \left\{ 
\omega \in \Omega (M^6) ~|~ t^* (\omega)= \alpha (t) \omega~{\rm ~for~all~} t \in \T^n
\right\}\;.
\label{Omega a}
\)
Correspondingly, 
we write a differential form in components as 
\(
\omega = \sum_{\alpha \in \hat{\T^n}} 
\omega_\alpha \;, \qquad \omega_\alpha \in \Omega (M^6)_\alpha\;.
\label{Exp w}
\)
Then the wedge product on components takes the form 
\(
\left(\omega \wedge \eta \right)_\alpha = \sum_{\alpha_1 + \alpha_2= \alpha} \left(
\omega_{\alpha_1} \wedge \eta_{\alpha_2}
 \right)\;.
\label{wedge}
\)
We then define the components of the deformed wedge product $\wedge_\theta$ to be 
\(
 \left(\omega \wedge_\theta \eta\right)_\alpha
:= \sum_{\alpha_1 + \alpha_2=\alpha} \omega_{\alpha_1} 
\wedge \eta_{\alpha_2}\, \theta (\alpha_1, \alpha_2)\;.
\)
The deformed wedge product $\wedge_\theta$ is no longer 
skew-symmetric in general, as we have that 
$\theta (\alpha_1, \alpha_2) = \overline{\theta (\alpha_2, \alpha_1) }$, i.e. 
$\theta$ is a phase. 

\vspace{3mm}
The action of the de Rham differential on the deformed wedge product is given by
\begin{eqnarray}
d( \omega \wedge_\theta \eta)_\alpha &=&
\sum_{\alpha_1 + \alpha_2 = \alpha} d(\omega_{\alpha_1} \wedge \eta_{\alpha_2}) \theta (\alpha_1, \alpha_2)
\nonumber\\
&=& \sum_{\alpha_1 + \alpha_2 = \alpha}
\left( d \omega_{\alpha_1} \wedge \eta_{\alpha_2} + (-1)^{{\rm deg}(\omega)} 
\omega_{\alpha_1} \wedge d \eta_{\alpha_2} \right)  \theta (\alpha_1, \alpha_2)
\nonumber\\
&=&
(d \omega \wedge_\theta \eta)_\alpha + (-1)^{{\rm deg}(\omega)}  (\omega \wedge_\theta d \eta)_\alpha\;,
\end{eqnarray}
hence
\begin{equation}
d( \omega \wedge_\theta \eta) = d \omega \wedge_\theta \eta + (-1)^{{\rm deg}(\omega)}  \omega \wedge_\theta d \eta\;.
\end{equation} 
Therefore the deformed wedge product $ \wedge_\theta$ induces a product on de Rham cohomology
\footnote{which still has the same classical definition since the differential and the space of forms 
have not changed.}
$H^\bullet(M^6)$. We will next compare this product structure with the product structure determined by the
standard wedge product. \\

Let $[0,1] \ni t \mapsto \theta_t \in Z^2(\widehat \T^n, U(1))$ be a 1-parameter family of cocycles. Then 
we get a homotopy $[0,1] \ni t \mapsto  \wedge_{\theta_t}$ of wedge products on cohomology,  and a standard
argument shows that $ \wedge_{\theta_0} =  \wedge_{\theta_1}$ on de Rham cohomology. Such homotopies are 
obtained by choosing $\xi \in Z^2(\widehat \T^n, \RR)$ and considering the homotopy 
$[0,1] \ni t \mapsto \theta_t = \exp(2\pi i t \xi)\, \theta \in Z^2(\widehat \T^n, {U}(1))$. By considering the long exact sequence 
in cohomology associated to the exact sequence of coefficients
$$
1\to \ZZ \to \RR \to { U(1)} \to 1
$$
and, noting that the torsion subgroup of $H^3(\widehat \T^n, \ZZ)$ is trivial, we conclude that the deformed
wedge product induces the same product structure on de Rham cohomology as does the usual wedge product.

Define the {\em  deformed algebra of differential forms} to be $\Omega^p (M_\theta^6) = \left(\Omega^p (M^6), \wedge_\theta\right)$.

%%%%%%%%%%%%%%
\subsection{Deformed Hodge star operator}
%%%%%%%%%%%%%
We will, as before, consider the worldvolume $M^6$ with a Riemannian metric $g$
and with an action of a torus $\T^n$ by isometries. 
 Considering isospectral deformations,
in which the metric is unchanged, the deformed Hodge star operator  
\(
*_\theta : \Omega^p (M_\theta^6) \to \Omega^{6-p}(M_\theta^6)
\)
is defined, as in \cite{LS}, by the commutative diagram 
\(
\xymatrix{
\Omega^p(M^6) 
\ar[rr]^-{*} 
\ar[d]^{L_\theta} 
&&
\Omega^{6-p} (M^6) 
\ar[d]^{L_\theta}
\\
\Omega^p(M_\theta^6)  
\ar[rr]^-{*_\theta}
&&
\Omega^{6-p} (M_\theta^6) \;. 
}
\)
That is, we define the deformed Hodge star operator on $H\in \Omega^p(M_\theta^6) $ by 
\(
*_\theta H=L_\theta * L_\theta^{-1} (H)\;.
\)
This will be one of the 
ingredients in building the action and the equations of motion.

\paragraph{The inner product on $\Omega(M_\theta^6)$ and the kinetic action functional.} 
The inner product, for two $p$-forms $H, H' \in \Omega^p(M_\theta^6)$, is 
\(
(H, H')_\theta= \ncint *_\theta (H^\dagger \wedge_\theta*_\theta H')
\label{inn}
\)
since $*_\theta (H^\dagger \wedge_\theta*_\theta H')\in C^\infty (M_\theta^6)$. 
Here $H^\dagger$ is the adjoint operator corresponding to the 
operator form of $H$. 
%
%\vspace{3mm}
%We can view our action as a pairing or inner product as follows. 
%\(
%(H_3^D, H_3^{D})_D=\ncint H_3^{D*} H_3^D
%\)
%\vspace{3mm}
We now consider how the noncommutative wedge product $\wedge_\theta$ and the 
noncommutative Hodge star operator $*_\theta$ work together. 
Using the above notions and definitions we build
 the expression that appears in the proposed action \eqref{Eq action}, that is
\(
H_3 \wedge_\theta *_\theta H_3\;.
%H_3 \wedge *(L_\theta (H_3)) \theta (3,3)
\)
An alternate expression for the inner product \eqref{inn}, is given by the 
noncommutative integral 
\(
 (H, H')_\theta = \displaystyle\int_{M^6_\theta}
  H^\dagger \wedge_\theta *_\theta H'\,.
  \label{inn2}
\)
It is easy to see the following symmetries of the inner product 
\(
 (H, H')_\theta  =  \overline{(H', H)_\theta}\;, \qquad  (aH, bH')_\theta
= \bar a b (H, H')_\theta\;.
\)
The nondegeneracy is explained in  \cite{BV} for example.
\footnote{in the Yang-Mills case, but the formulation is general.}
Either of the two expressions, \eqref{inn} or \eqref{inn2}, 
 can be taken to be
  the kinetic term of the H-field. Note that for $H=H'$ this implies that the 
inner product is real.
This is also a generalization/analog of the nonabelian Yang-Mills description in four dimensions 
in \cite{LS} to abelian gerbe theory in six dimensions.

%%%%%%%%%
\subsection{Torus bundles and strict deformation quantization}
%%%%%%%%%

The fivebrane worldvolume theory can be considered on tori as well as
on torus bundles. Different fiber dimensions capture different 
physical aspects of the theory. As an application of our construction, 
we discuss here the example the case of a principal 2-torus bundle, which is also physically
relevant.  

\medskip
Consider fivebrane worldvolume $M^6$ 
as the total space of a principal 2-torus bundle over an oriented 4-dimensional manifold $X^4$.
As before, we strictly deform 
quantize with respect to the 2-torus action to get a noncommutative principal torus bundle (NCTP) with 
total space $M^6_\theta$ and
constant deformation parameter $\theta$, cf. \cite{HM, HM2}.   NCTP bundles occur in the study 
of T-duality in a background flux \cite{T1,T2,T3} in string theory, and was first described 
in terms of strict deformation quantization in \cite{HM, HM2}.

 \vspace{3mm}
 We start by recalling the commutative case, from \cite{Ve, D, W-6}. 
The classical Hodge star operator $*: \Omega^p(M^6) \to \Omega^{6-p}(M^6)$
depends only on the conformal class of the metric on $M^6$. 
Classically, the equations 
\(
d^\dagger H_3= \pm *d* H_3=0\;, \qquad dH_3=0
\label{EOMclass}
\)
are therefore also conformally invariant. The passage to the quantum theory preserves this 
property since the theory is linear \cite{W-6}. 

\medskip
Consider the worldvolume as the product $M^6=X^4 \times \T^2$ with the 
product conformal structure. The reduction of the theory on $\T^2$ results in 
an induced 4-dimensional theory  which depends on the conformal structure
of $\T^2$ up to isomorphism. The latter is determined by a choice of a 
point $\tau$ in the upper half plane modulo the action of the modular 
group $SL(2, \Z)$. With complex coordinates $z=x + iy$ the ansatz
for the H-field is 
\(
H_3= F_2 \wedge dx + *_4 F_2 \wedge dy\;,
\label{ansatz}
\)
where $F_2$ is a two-form pulled back from $X^4$ 
to $M^6$ and $*_4$ is the Hodge star operator on $X^4$. 
Classically, the Bianchi identity (or equation of motion with self-duality condition)
$dH_3=0$ gives rise to Maxwell's equations $dF_2 = d*_4 F_2=0$.

\medskip
We propose a description 
in terms of the noncommutative deformation. 
Equation \eqref{EOM3} is then conformally invariant in the noncommutative sense. 
This is an important ingredient in viewing (a limit of) the theory as an (exotic)
(2,0) conformal field theory. 
In the special case when 
$M^6$ is the trivial bundle  $X^4 \times \T^2$, we have that $M^6_\theta$ is just  
$X^4 \times \T^2_\theta$.
A description of the 
noncommutative conformal structures on noncommutative 2-torus $\T^2_\theta$
is given in
\cite{CM2}. Including a Weyl factor makes the 
flat geometry on $\T^2_\theta$ into a curved geometry.

\medskip
We now consider what happens upon dimensional reduction, 
providing a generalization in the sense that the ansatz \eqref{ansatz} 
is replaced with a noncommutative counterpart. Generally,
 to account for nontriviality of the torus bundle, the ansatz, relating 
$H_3$ to $F_2$ and its dual, is replaced by integration over the fiber. 
The first operation to deform is  the wedge product $\wedge$, which would be replaced
with $\wedge_\theta$. This then leads to a noncommutative version of 
dimensional reduction, whereby the resulting equations obtained from the Bianchi 
identity $dH_3=0$ are the same commutative Yang-Mills equations 
$dF_2=0=d*_4F_2$.

\medskip
Note that if, in addition, we deform the Hodge operator $*$ and replace it with their noncommutative
counterpart $*_\theta$, then with the new ansatz, the Bianchi identity $dH_3=0$ leads to 
the noncommutative equations $dF_2=0$ and $d*_\theta F_2=0$, which are the Bianchi 
identity and equation of motion of noncommutative Yang-Mills theory. 
It would be interesting to discuss the extension to this most general case, 
which requires $X^4$ to be noncommutative as well.

\medskip
Note that other torus bundles of other fiber dimensions are possible.
In fact, one can go all the way and consider the 
worldvolume as a six-torus. This is considered for example in 
\cite{DN, He}. We can view this as a limiting 
case of a 6-torus fiber with a point as a parameter space, and the
corresponding deformation is covered by our discussion.

%%%%%%%%%%%%%
\section{The partition function of the model}
%%%%%%%%%%%%%
\label{Sec pf}
In this section, as an application, we consider  a variation on the M5-brane worldvolume theory 
as a model for studying S-duality. The main point  will be that, to a large
extent, we are able to have an analogous discussion on this model of 
deformed 6-dimensional higher abelian gerbe theory as has been done
in the undeformed 4-dimensional abelian Yang-Mills case. 

\subsection{Evaluation of the partition function} 
%%%%%%%%%%

we will outline the evaluation of the partition function. 
We have found the discussion in \cite{PS} and \cite{EN} in three and four dimensional abelian Yang-Mills
theory, respectively, to be very useful. Our discussion is a generalization to six dimensions 
and an extension to include deformations. We will start with the rational case and then 
work globally, making use of the moduli spaces associated to gerbes introduced in \cite{MR}. 

\paragraph{Deformed Hodge theory.}
In order to proceed, we argue that Hodge theory in general, and the 
Hodge theorem in particular, continues to hold in the noncommutative setting. 
Consider the following observations.\\

1) The cohomology of the deformed de Rham complex does not
depend on the deformation, since the differential is the same and the
(wedge) product is not used in the definition of cohomology.\\

2) Since the deformation is isometric, the deformed Laplacian is isospectral
to the undeformed one, so in particular there is an isomorphism of the 
respective zero eigenspaces.\\

Using these two facts, it follows that Hodge theory in the deformed case is a consequence of 
standard Hodge theory (in the undeformed case).

%%%%%%%%%%%%%%%%%%%%%%%%%%%
\paragraph{Rationally with no topological terms.}
%%%%%%%%%%%%%%%%%%%%%%%%%
We start with the theory described by the action \eqref{action for pf} with no topological term, i.e. with the
parameter $\Theta$ set to zero. This then reduces to the Yang-Mills gerbe theory 
described by \cite{MR}. Furthermore, we will first work at the level of 
differential forms, where the cohomology class of the gerbe is trivial. 
With $\mathbb{G}$ the gauge group, i.e. the group of gauge transformations, the partition function is then given by
\begin{eqnarray}
Z(M_\theta^6, e)&=&\frac{1}{{\rm Vol}(\mathbb{G})} \int_{\Omega^2_\theta(M_\theta^6)} 
DB e^{\tfrac{i}{2e^2} \int_{M_\theta^6} H_B \wedge_\theta \ast_\theta H_B}
\nonumber\\
&=& 
\frac{1}{{\rm Vol}(\mathbb{G})} \int_{\Omega^2_\theta(M_\theta^6)} 
DB e^{\tfrac{i}{2e^2} \langle B, d_2 d_2^\dagger  B \rangle}\;,
\end{eqnarray}
where $d_2$ is the exterior derivative acting on 2-forms $\Omega^2_\theta(M_\theta^6)$,
and $d_2^\dagger$ is the deformed adjoint $\ast_\theta d_2 \ast_\theta$. We need the labeling 
on the operators in order to distinguish the appearance of various guises of this operator in 
the partition function. 
In going from the first line to the second we have used properties of this derivative
on deformed differential forms we derived in the previous section . 

\medskip
Decomposing the space as $\Omega^2_\theta(M_\theta^6)=\ker d_2 \oplus \big(\ker d_2 \big)^\perp$,
then we form the space obtained by the action of the gauge group $\mathbb{G}$ as
\begin{eqnarray}
[B]&=& {\mathbb G} \cdot B
\nonumber\\
&=& (\ker d_2) \cdot B
\nonumber\\
&=& \{B + B'\;,  ~ {\rm where}~ d_2 B'=0\}\;.
\label{global trans}
\end{eqnarray}
Generally, one can write the partition function as an integral over the space of fields ${\cal C}$, or
as an integral over the moduli space of fields ${\cal M}_B$, which is the space of fields modulo gauge transformations
${\cal C}/{\mathbb{G}}$.
The two are related by the formula
\(
Z(M, \lambda)=\frac{1}{{\rm Vol}(\mathbb{G})} \int_{\cal C}  DB~e^{-i \lambda S(B)}
=\frac{1}{{\rm Vol}(\mathbb{G})} \int_{{\cal C}/\mathbb{G}}  D[B]~{\rm Vol}([B])~e^{-i \lambda S([B])}\;.
\)
With this, factoring the volume of `gauge group' we get
 \begin{eqnarray}
Z(M_\theta^6, e)&=& 
\frac{{\rm Vol}(\ker d_2)}{
{{\rm Vol}(\mathbb{G})}} 
 \int_{\big(\ker d_2 \big)^\perp} D[B]~ 
 e^{\tfrac{i}{2e^2} \langle B, d_2 d_2^\dagger  B \rangle}
 \nonumber\\
 &=& 
%\frac{{\rm Vol}(\ker d_2)}{
%{{\rm Vol}(\mathbb{G})}}  
{\rm det}{}'  \Big(\frac{-i}{4\pi e^2} d_2 d_2^\dagger \Big)^{-\tfrac{1}{2}}\;.
\end{eqnarray}
Both terms are infinite, and so a regularization is needed. We will use $\zeta$-function 
regularization. 

\paragraph{Interpretation of the terms.} 
We have local gauge transformations of the B-field $B \mapsto B_2 + dA$, where 
\footnote{from here on we will drop the subscript and write $\Omega$ for $\Omega_\theta$.}
$A_1 \in \Omega^1(M_\theta^6)$. This is a local version of the global transformations 
in \eqref{global trans}. The stabilizer of the group of such 
transformations is flat connections 
\(
{\cal F}={\rm Hom}\big(\pi_1(M_\theta^6), U(1)\big)/U(1)=\left\{A \in \Omega^1(M_\theta^6)~|~ d_1 A=0 \right\}\;,
\)
where $d_1$ is the exterior derivative acting on 1-forms. If the fundamental group is abelian then, by 
the Hurewicz theorem,  $\pi_1(M_\theta^6)_{\rm ab}\cong H_1(M_\theta^6)$. Furthermore, since the conjugation 
action of $U(1)$ is trivial, then in this case 
\(
{\cal F}={\rm Hom}\big(H_1(M_\theta^6), \R/\Z\big)\;.
\)
Now the integration can be taken over this quotient space. 

\medskip
In order to choose a representative from each equivalence class $[B]$, we 
impose a gauge condition. A convenient one is the degree two analog of the Lorentz gauge, 
i.e. $d_1^\dagger B=0$. The gauge-fixing condition can be imposed via a delta function. 
This is the Faddeev-Popov ghost determinant $\det(d_2d_2^\dagger)$. Now, as the case
of other gauge theories, the operator $d_1$ acting on $A$ itself has zero modes. 
This implies that we need a ghost-for-ghost determinant, which amounts to dividing by the volume 
of the stabilizer ${\cal F}$ at each point.
% assuming a normalization of ${\rm Vol}({\cal G})=1$. 
This in turn needs a gauge-fixing  condition, which now in this degree
 literally looks like the one in abelian Yang-Mills theory, say the Lorentz gauge $d_0^\dagger A=0$. 

\paragraph{Steps in the evaluation of the partition function.} 
We now go through the following steps:
\begin{enumerate}
\item We replace ${\rm Vol}(\ker d_2)$ by ${\rm Vol}(\ker d_1)$. 

\item ${\rm Vol}(\ker d_1)$ can be evaluated as follows. 
The projection map $\ker d_i \to H_{\rm dR}^i(M_\theta^6)$ indices an isomorphism 
$\phi_i: {\rm Harm}^i(M_\theta^6) \buildrel{\cong}\over{\longrightarrow} H^i_{\rm dR}(M^6_\theta)$.
This follows from the similar isomorphism in the undeformed case. Then, by 
change of variables, we have 
${\rm Vol}({\rm Harm}^i(M_\theta^6))=| \det \phi_i|^{-1} {\rm Vol}(H_{\rm dR}^6(M_\theta^6))$.
This then gives that the volume of the stabilizer is 
${\rm Vol}(\ker d_1)=\det (\phi_1^\dagger \phi_1)^{\tfrac{1}{2}}\cdot{\rm Vol}({\rm Harm}^i(M_\theta^6))$.

\item The ghost-for-ghost determinant $\det (\phi_1^\dagger \phi_1)^{\tfrac{1}{2}}$ extracts the zero modes from the 
Faddeev-Popov determinant. This is given by inverse volume of the manifold, as in the classical case (e.g. \cite{PS}), 
i.e $\det (\phi_1^\dagger \phi_1)=\big( {\rm Vol}(M_\theta^6)\big)^{-1}$, since the metric properties are 
unchanged in the spectral deformation.  

\item The complex scaling factor $\tfrac{-i}{4\pi e^2}$ can be factored out at the expense of introducing the 
$\eta$-function and the $\zeta$-function of the operator $d_2^\dagger d_2$, 
\(
\det\left(\tfrac{-i}{4\pi e^2} d_2^\dagger d_2 \right)^{-\tfrac{1}{2}}=
e^{-i \pi \eta(d_2^\dagger d_2)} \left(\tfrac{-i}{4\pi e^2}\right)^{-\tfrac{1}{2}\zeta (d_2^\dagger d_2)}
{\rm det}'(d_2^\dagger d_2)^{-\tfrac{1}{2}}\;,
\)
where $\eta(P)=\eta (0, P)$ and $\zeta(P)=\zeta(0, P)$ are the analytic continuations of the 
operator eta-function and zeta-function for a differential operator $P$ of Laplace type.  
Now the zeta function of our operator is given by 
$\zeta (d_2^\dagger d_2)=\sum_{j=0}^2 (-1)^j \zeta (\Delta_j)$, where
$\Delta_j$ is the Hodge Laplacian on $j$-forms. The zeta function of the latter in turn is 
given by  
\(
\zeta (\Delta_p)= - \dim H^p_{dR} (M_\theta^6) + {\rm corrections}.
\label{corr}
\)
The first term can be replaced by the corresponding Betti numbers, as follows 
via the deformed version of the de Rham theorem. The second term will be considered 
in the next section when investigating modularity.

\end{enumerate}

Overall, with the above contributions, the partition function finally takes the form 
\(
Z(M_\theta^6, e)= e^{-i \pi \eta(d_2^\dagger d_2)}
 \left(\tfrac{-i}{4\pi e^2}\right)^{-\tfrac{1}{2}(b_2 - b_1 + b_0)}
 \frac{{\rm Vol}(M_\theta^6)^{\tfrac{1}{2}}}{{\rm Vol}({\rm Harm}^1(M_\theta^6))}
 \frac{\det'(d_1^\dagger d_1)^{\tfrac{1}{2}}}{\det'(d_2^\dagger d_2)^{\tfrac{1}{2}}}
\;,
\label{PF explicit}
\) 
 up to factors involving asymptotic expansions of the heat kernel. 

\medskip
Adding the purely topological 
term $\int_{M_\theta^6} H_B \wedge_\theta H_B$ 
would   
amount to a numerical correction to the 
partition function. This is analogous to the instanton term 
in Yang-Mills theory.

%%%%%%%%%%%%%%%
%\subsection{Globally, with instantons and torsion}
%%%%%%%%%%%%%%%

\medskip
We now recast the above discussion in a more global context. 
Let ${\cal G}$ be an abelian gerbe on $M_\theta^6$. 
Isomorphism classes are classified by the Dixmier-Douady class $DD ({\cal G})\in H^3(M_\theta^6; \Z)$. 
Then $B$ can be viewed as a 2-connection with curvature $H_B$. 
Fix a gerbe ${\cal G}_0$ over $M_\theta^6$ and a 2-connection $B_0$ on it such that 
$H_B(B_0)\in \Omega^3(M_\theta^6)$. The space of 2-connections on $\cal{G}$ relative to 
$B_0$ is 
\(
{\cal B}(B_0):=\left\{B~|~B=B_0 + b ~{\rm with}~ b \in \Omega^2(M_\theta^6)  \right\}\;.
\)
If ${\cal G}_0$ is the trivial gerbe and $B_0$ is the trivial 2-connection then 
${\cal B}(B_0)\cong \Omega^2(M_\theta^6)$. 
\footnote{Note that in order to impose integrability and introduce topology, we can take 
$H_3(B_0)\in L^2(M_\theta^6, \Lambda^3M_\theta^6)$
and then ${\cal B}(B_0) \cong L^2(M_\theta^6; \Lambda^6 M_\theta^6)$, analogously to the 
Yang-Mills case in \cite{EN}.}
The space ${\cal{B}}(B_0)$ is acted upon by gauge transformations 
$\mathbb{G}$ and the orbit space of gauge equivalence classes with its quotient 
topology is ${\cal O}(B_0)={\cal B}(B_0)/\mathbb{G}$. 

\medskip
The moduli space of the solutions to the equation of motion of the system 
can be described explicitly in the classical case \cite{MR}. Our current discussion is
a straightforward extension to the deformed case. 

\begin{itemize}
\item Let ${\cal M}_B$
denote the moduli space of gauge inequivalent solutions of the deformed equation of motion
\eqref{EOM3}
associated to a bundle gerbe 
$\cal{G}$ with connection $B$ over $M_\theta^6$. Then ${\cal M}_B$ is 
diffeomorphic to the torus $T^{b_2(M)}$ of dimension equal to the second Betti number of $M_\theta^6$. 
That is, we have 
\(
{\cal M}_B = H^2(M_\theta^6; \R)/ H^2(M_\theta^6; \Z) \cong T^{b_2(M)}\;,
\)
where the quotient group can be viewed as the gauge group in this case. 

\item
The space ${\cal B}$ of all bundle gerbe connections on $\cal{G}$ over $M_\theta^6$
is an affine space associated with the vector space $\Omega^1({\cal{G}})/\pi^*(M_\theta^6)$, where
$\pi$ is the projection from the gerbe to $M_\theta^6$. Then, as in \cite{MR}, the total moduli space 
$\cal{M}=\bigcup_{B \in {\cal B}} \cal{M}_B$ is diffeomorphic to a torus $T^{b_2(M)}$-bundle 
over the affine space ${\cal B}$. 

\item We can now include torsion to account for gerbes with torsion curvatures. These have 
the virtue of being described by finite-dimensional bundles.  
We start with the decomposition 
\begin{eqnarray}
H^2(M_\theta^6; \Z) &\cong & {\rm Free}(H^2(M_\theta^6; \Z)) \times {\rm Tor}(H^2(M_\theta^6; \Z))
\nonumber\\
&\cong & \Z^{b_2(M)} \times {\rm Tor}(H^2(M_\theta^6; \Z))\;.
\end{eqnarray} 
The the space of gauge equivalence classes of flat gerbes on $M_\theta^6$ is 
\(
\widehat{\cal M} \cong T^{b_2(M)} \times {\rm Tor}(H^2(M_\theta^6; \Z))\;.
\)
\end{itemize}

We now describe the effect on the corresponding partition function, generalizing the above
 discussion in the
rational case to include  
contributions from the torsion part. Gerbes with line bundles $L$ such that $c_1(L) \in 
{\rm Tor}(H^2(M_\theta^6; \Z))$ are flat, which implies that the curvature $H_B$ of the 
corresponding bundle gerbe ${\cal G}$ is trivial. Consequently, the action vanishes along 
these. The contribution will then be simply a numerical factor given by the 
rank of the torsion part, i.e. $|{\rm Tor}(H^2(M_\theta^6; \Z))|$.

\medskip
The partition function over the full moduli space is 
\bea
Z(M_\theta^6, \tau)&=& \sum_{[B_0]\in \widehat{\cal M}} \int_{[B]\in \cal{M}}
e^{-S(B, \tau)}
\nonumber\\
&=&
\frac{1}{{\rm Vol}(H^2(M_\theta^6; \Z))} 
\int_{B \in H^2(M_\theta^6; \Z)} D[B] e^{-S(B, \tau)}\;.
\eea
By the transformations $H_B \mapsto H_B + db$, we arrive at calculations that
are similar to the rational case, but now with $b$ here in place of $B$ there. 
%So we get factors like $\langle b, \Delta_2 b\rangle$ etc. 
The remaining discussion, including the effect of the full moduli space and that of torsion, 
 is then a straightforward extension of expression 
\eqref{PF explicit}, which we will leave for the reader to verify.

\medskip
We have set up the partition function in analogy to the 4-dimensional 
(abelian) Yang-Mills case. 
A detailed analysis
% of the integral in \eqref{eq sum} 
requires understanding of the geometry
and topology of the moduli space ${\cal M}_\cG$, in particular the renormalized volume and 
integrals, and also an extensive 
discussion, adapting that of the seminal work of Verlinde \cite{Ve}, Witten \cite{Wi2}, 
Vafa-Witten \cite{VW} to the  case of 
gerbes, going beyond the scope of this letter. 
%What 
%we focus on here is the modularity of the partition function and not its 
%particular expression, and the value of these integrals presumably will not spoil 
%modularity as in the case considered in the above cited papers. 
Of course we expect that the full evaluation of $Z(M^6_\theta; \tau)$ 
might lead to connections to geometric and topological invariants of 
$M_\theta^6$, as in the commutative case (see \cite{W-6, W-knot, S1, S2}).

%%%%%%%%%%%%%%%%%%%%%%%%%%
\subsection{Modularity of the partition function}
% of the model}
%%%%%%%%%%%%%%

We will consider modularity of the partition function, as an extension to 
gerbes and to deformed spaces
of the Maxwell case in four dimensions. For completeness and ease of 
comparison we first recall the latter.

\paragraph{The 4d partition function as a modular form.} 
There are two ingredients in the 4-dimensional abelian case \cite{Wi2, Ve}. First, the modular 
transformations of ${\rm Im}(\tau)$ are as follows. ${\rm Im}(\tau)$ is invariant under a $T$-transformation, 
i.e. ${\rm Im}(\tau+1)={\rm Im}(\tau)$, and transforms under an $S$-transformation as 
${\rm Im}\big(\tfrac{-1}{\tau}\big)=\frac{1}{\tau \overline{\tau}}{\rm Im}(\tau)$. The two transformations 
together imply that ${\rm Im}(\tau)$ is a modular form of weights $(-1, -1)$. 
Second, the Narain-Seigel $\theta$-function of the lattice of signature $(b_2^+, b_2^-)$ is a modular form
of weights $(\tfrac{1}{2}b_2^+, \tfrac{1}{2}b_2^-)$. Combining the two gives that the partition function of 
a free abelian gauge theory on a 4-manifold $X^4$ is a modular form of 
weights $\tfrac{1}{4}(\chi - \sigma, \chi + \sigma)$, where 
$\chi=\chi (X^4)$ and $\sigma=\sigma(X^4)$ are the Euler characteristic and the signature, respectively, of 
$X^4$. These combinations arise because $\chi=2b_0 - 2b_1 + b_2^+ + b_2^-$ and 
$\sigma= b_2^+ - b_2^-$.
%with assumption $b_0=b_1$.

\paragraph{The general form of the partition function for the abelian gerbe theory.}
We now go back to our problem and consider the action as in 
\eqref{Eq action} 
 but drop the term involving $C_6$ for simplicity 
because it does not 
affect the modularity argument. 
Assuming as before that 
3-curvature is self-adjoint $H_B^\dagger = H_B$ with respect to the deformed inner product,
the action is
\(
S_\theta(B; \tau)= \frac{1}{2e^2}\int_{M_\theta^6} H_B \wedge_\theta *_\theta H_B + 
\frac{i \Theta}{2} \int_{M_\theta^6} H_B \wedge_\theta  H_B\;,
\label{action for pf}
\)
as in the case for analyzing the partition function in the Yang-Mills case. 
The second term in the action \eqref{action for pf} is a topological invariant,
involving the Dixmier-Douady classes of the underlying gerbe. This then gives
that the integrand $e^{-S_\theta (B, \tau)}$ of the partition function is 
always invariant under the transformation $\Theta \mapsto  \Theta + 4\pi$.

\medskip
For purposes of modularity of the partition function rather than for explicit 
evaluation of it, one can in principle take the fields in the theory to be either
the curvatures or the connections, as is also done in the four-dimensional 
abelian undeformed case. Choosing curvatures will amount to summing over degree three 
cohomology classes in place of degree two classes for the connections.
As the latter was used explicitly in calculating the partition function in the previous section, 
for completeness we will also
highlight the former in discussing modularity int his section.

\medskip
The partition function is set as above, and similarly to the Maxwell case in four dimensions, as 
\(
Z(M^6_\theta; \tau)= \sum_{[\cal{G}]} \frac{1}{\vol(\cM_{\cG})} \int_{{\cal M}_{\cG}} DB\, e^{-S_\theta (B, \tau)}\;, 
\)
where the sum is over equivalence classes of gerbes $\cal{G}$ on $M_\theta^6$ and the integral is over 
$\cal{M}_{\cG}$, the moduli space of B-fields on $\cG$, with a measure $DB$.

Using deformed Hodge theory, we proceed as in the abelian Yang-Mills case, by decomposing 
the B-field as 
\(
B= B_0 + B_h^{\cal{G}}\;,
\)
where $B_0$ is a 2-connection on the trivial gerbe $\cG_0$ which is a global 2-form on $M_\theta^6$
and $B_h^{\cal{G}}$ is any 2-connection on the gerbe $\cal{G}$ 
having harmonic curvature $H_h^{\cal G}$. 
The partition function then takes the form 
\(
Z(M^6_\theta; \tau)= \frac{1}{\vol(\cM_{\cG_0})}\int_{{{\cal M}}_{\cG_0}} DB_0 \, e^{-S_\theta (B_0, \tau)}\sum_{[\cG] \in H^3(M^6_\theta; \Z)} e^{-S_\theta(B_h^{\cal G})}\;.  
\label{eq sum}
\)
We now consider the sum inside the expression for the partition function. 
On the lattice $H^3(M^6_\theta; \Z)$ we have two quadratic forms,
both in a sense intrinsically noncommutative; for 
$\omega$ a $\theta$-harmonic (i.e. noncommutative-harmonic) 3-form, these are
\begin{eqnarray}
q_1&:=&\int_{M_\theta^6} \omega \wedge_\theta \omega\;,
\nonumber\\
q_2&:=&\int_{M_\theta^6} \omega \wedge_\theta *_\theta \omega\;.
\label{new quad}
\end{eqnarray}
The first quadratic form $q_1$ is intrinsically noncommutative, as it vanishes
identically in the commutative case, with or without self-duality. 
The second quadratic form vanishes in the commutative self-dual case,
as for the M5-brane theory (without deformation). However, it is also non-identically vanishing in the 
noncommutative setting, even when self-duality is imposed. 

\medskip
The quadratic form $q_1$ is indefinite in general with signature, as in the 
degree two Yang-Mills case, $\sigma(q_1)=(b_3^+, b_3^-)$, 
where $b_3^\pm$ are the dimensions of the selfdual and ant-selfdual 
$\theta$-harmonic 3-forms, respectively, 
\(
{\rm Harm}_\theta^{3, +}=\left\{ 
H_3^\pm \in \Omega^3(M_\theta^6) ~\vert~ \Delta_\theta H_3^{\pm}=0, 
~ H_3^\pm=H_3 \pm *_\theta H_3
\right\}\;,
\)
where $\Delta_\theta$ is the noncommutative Hodge Laplacian. 

\medskip
We now get back to the expression of the partition function. Taking 
$\omega=\tfrac{1}{2\pi}H_3^{\cal G}$, the sum over gerbes in 
\eqref{eq sum} becomes 
\(
\sum_{\omega \in H^3(M_\theta^6; \Z)} \exp \left(  
-\tfrac{4\pi}{e^2}q_2 + i \tfrac{\Theta}{2}q_1
\right)\;.
\)
With the  identification $\tau=\tfrac{\Theta}{2\pi} + \tfrac{4\pi i}{e^2}$, this
sum is a modular form with holomorphic/anti-holomorphic weights 
$(\tfrac{1}{2}b_3^+, \tfrac{1}{2}b_3^-)$. 

%\medskip
%\fbox{The above two paragraphs require some precision} 

\medskip
 Let us consider this more carefully and in more detail. 
The number of zero modes of the Hodge Laplacian on $p$-forms 
coincides with the $p$-th Betti number $b_p$. 
Since we are in an even dimension not divisible by 4, the middle Betti number 
$b_3$ is always even. In this case,
the gerbe contributes $b_2 - b_1 + 1$, while the $\theta$-function 
contributes $b_3$. As cancellation is not possible in general in this case,  
there is a modular anomaly, i.e. the partition function will 
transform as a modular form of nonzero weights. A priori, and under conditions on lower modes,
these weights are expected to be of the form $(\tfrac{1}{2} b_3^+, \tfrac{1}{2}b_3^-)$. 
Furthermore, one can ask whether these can be written in terms of local indices or topological 
invariants, as was the 4-dimensional case.

\paragraph{Example 1: The six-torus $\T^6$.} This is an example where all the Betti numbers are present. 
By the K\"unneth theorem, the Poincar\'e polynomial of 
$\T^6$ is $P(\T^6)=(1 + x)^6$, from which the Betti numbers are read off as the binomial coefficients,
i.e. $b_0=1$, $b_1=6$, $b_2=15$, $b_3=20$. Then the Euler characteristic is $\chi(\T^6)=0$. 
The gerbe contributes $b_2-b_1+1=10$ zero modes, while the $\theta$-function a priori gives
$b_3=20$. However, taking into account self-duality the latter instead gives  a matching
$\tfrac{1}{2}b_3=10$. Note that the modularity of the partition function of the M5-brane is 
demonstrated in \cite{DN}. As our deformations are isospectral, the same result holds
for the M5-brane with worldvolume the noncommutative  space $\T_\theta^6$.

\paragraph{Example 2: (Product of) Lie groups $G$.} This is an example in which only the 
third Betti number is 
relevant. Let $G$ be a compact semi-simple Lie group.
Such a group is 2-connected because being semi-simple implies that it is simply connected, and
being a Lie group implies that $\pi_2(G)=0$. So $b_1=0=b_2$. The third Betti number is
$b_3=r$, the number of nonabelian summands i the decomposition of the Lie algebra of 
$G$ into simple algebras. In this case the gerbe can be taken as the basic gerbe on $G$
whose zero modes are characterized by $b_3$.
Likewise, the $\theta$-function is characterized by $b_3$. Thus, viewing the 
field as a degree three class, one can arrange for a cancellation, hence obtaining 
a modular form. The main example in this case is the product of two spheres and 
quotients thereof.  Again, due to the deformation being isospectral the results
extend to the noncommutative case, e.g. to $M_\theta^6=S^3_\theta \times S^3_\theta$ with 
the natural torus action.

\paragraph{Modular weights via topological invariants?}
We would like to investigate this in the noncommutative
setting. We have argued above that Hodge theory continues to hold and that, since 
spectra are preserved, the Betti numbers are also preserved. While our answers  to the above 
question will not be complete, we will be provide some arguments in the general case 
and illustrate with some examples. Furthermore, along the way
we will discover a new phenomenon due to noncommutativity.

\paragraph{1. The Euler characteristic.} 
We start by considering effects due to being in six dimension, still considering classical manifolds.
The Euler characteristic of $M^6$ is given by $\chi (M^6)= \sum_{i=0}^6 (-1)^i b_i$
which, by the duality $b_k=b_{6-k}$, becomes $\chi(M^6)=2b_0 - 2b_1 + 2b_2 -b_3$. Since
$b_3$ is even in six dimensions, then so is the Euler characteristic. This can be simplified if 
the lower Betti numbers either vanish or cancel each other out. 
In particular, if $M^6$ is 2-connected then $\chi(M^6)=2 - b_3$. It is a useful result of Smale that
a 2-connected compact 6-manifold is diffeomorphic to either the 6-sphere $S^6$ or the connected sum
of finite copies of the product $S^3 \times S^3$. For the former case we have $\chi (S^6)=2$ and for the 
latter case $\chi(\#_m S^3 \times S^3)=2-2m$, as $b_3=2m$ in this case. For $m=1$ gives 
$\chi (S^3 \times S^3)=0$. A main point here is that  all of this applies to when $M^6$ is replaced by its noncommutative
deformation $M^6_\theta$. 

\paragraph{2. The signature.} 
Classically, the signature of $M^6$ is given by $\sigma = {\rm sign} (M^6)=b_3^+ - b_3^-$, where $b_3^\pm$ are
the eigenvalues of the chirality operator. However, since we are in an even dimension which is 
not a multiple of 4, i.e. $n=6=2 \cdot 3$ and $3/4 \notin \Z$, we have that the quadratic form 
\(
q(u, v)= u \cup v= (-1)^{3^2} v \cup u = (-1)^{3^2} q(v, u)\;,  \qquad u, v \in H^3(M^6; \Z),
\)
corresponding to the intersection matrix is 
skew-symmetric. This implies vanishing signature $\sigma=0$ and so $b_3^+=b_3^-$. With this, 
$b_3=2b_3^+=2b_3^-$. When $M^6=\#_m (S^3 \times S^3)$, the intersection form
is isomorphic to the connected sum of $m$ copies of the standard skew-symmetric hyperbolic
form ${\cal H}(\Z)$ on the lattice $\Z \times \Z$. However, upon deformation, the cup product 
and in particular the wedge product are no longer skew-symmetric (cf. quadratic 
forms \eqref{new quad}) and 
so this signals that
the signature is not zero. Therefore, one of the byproducts of our analysis is that: {\it noncommutative
deformations also gives us the ability to 
define the signature operator in dimensions $n=4k+2$, which are classically vanishing}. 
It would be interesting to develop the theory of such signature operators further. We hope to take this
up elsewhere.

%%%%%%%%%%%%%%%%%%%%%%%%%
\subsection{Curvature corrections to modular weights}
%%%%%%%%%%%%%%%%%%%%%%%%%%

The partition function has a modular anomaly. 
We have a modular form of weights $({\rm wt}_1, {\rm wt_2})$, i.e. transforms as
\(
Z\Big(M_\theta^6, \frac{a\tau + b}{c\tau +d}\Big)= (c\tau + d)^{{\rm wt}_1} (c\overline{\tau} + d)^{{\rm wt}_2} Z(M_\theta^6, \tau)
\)
under an ${\rm SL}(2, \Z)$ transformation. 

\medskip
We will now consider curvature corrections ${\rm corr}_i$, $i=1,2$ to the modular weights 
\(
({\rm wt}_1, {\rm wt}_2) \mapsto ({\rm wt}_1 + {\rm corr}_1, {\rm wt}_2+ {\rm corr}_2)\;,
\)
which arise form the curvature corrections to the formula \eqref{corr}. We will assume that the construction of 
 heat kernels goes through for our isospectral noncommutative deformations. 
Then, with this assumption and in our case of an even dimension 
these take the form  (see e.g. \cite{Ros})
\(
\tfrac{1}{16\pi^2}\int_{M_\theta^6} {\rm tr}(u_p^6)dV\;,
\)
which is part of the expansion of the heat kernel, namely the sections 
$u_p^n \in C^\infty(M_\theta^6, {\rm End}(\Lambda^nM_\theta^6))$, $n=0, 1, \cdots$ 
appear in the coefficients of the short time asymptotic expansion of the heat kernel of the 
Laplacians on $p$-forms
\(
\sum_{\lambda \in {\rm spec}(\Delta_p)} e^{-\lambda t} 
~\sim~
\tfrac{1}{(4\pi t)^2} \sum_{n=0}^\infty \left(\int_{M_\theta^6} {\rm tr}(u_p^n) dV\right) t^{\tfrac{n}{2}} 
\qquad {\rm as}~ t \to 0\;. 
\)
These coefficients can be calculated in terms of the various curvatures of the 
manifold, which we assume extend to deformed case $(M_\theta^6, g)$.
These can be found explicitly in \cite{Gil}. Unlike the 4-dimensional case, where
these take elegant form in terms of curvature invariants, the 6-dimensional 
expression is both considerably more complicated and less elegant. 

%\medskip
%\fbox{Should we record these nevertheless?}
%

\medskip
We can consider the example of a 2-torus bundle $\T^2 \to M^6 \to X^4$. 
The fivebrane partition function on the undeformed 6-dimensional worldvolume $M^6$ has only a 
$\Z_2$ symmetry, but the strictly deformed
noncommutative manifold $\T^2_\theta \to M^6_\theta \to X^4$ 
has a fivebrane partition function with the full $SL(2, \Z)$ 
symmetry.
 
 \medskip
Overall, we have that our approach via non-commutative deformation evades problems with 
self-duality in general, and restores modular invariance of the partition function 
in the model discussed in this section. We hope that main features of these observations will  be of use in
further explorations and in other settings. We also hope that the noncommutative
formulation of the higher abelian gerbe theory on the M5-brane, in the sense discussed in earlier sections,  
will be useful in further investigations of that theory. 

\medskip
Note that the calculation of the partition function might be more tractable if lifted to 
eight dimensions, via the Chern-Simons construction, as in \cite{Wi, S2}. We leave this
for future investigation. 

%%%%%%%%%%%%%%%%%%%%%%%%%%%%%%%%%%%%%%%%%%%%%

 \end{document}